\documentclass[aps,prl,twocolumn,reprint,nofootinbib]{revtex4}

\usepackage[dvips]{graphicx}
\usepackage{url}
\usepackage{bm}
\usepackage{braket}
\usepackage{color}
\usepackage{comment}

\usepackage[dvipdfmx]{hyperref}
\hypersetup{
    colorlinks=true,
    citecolor=blue,
    linkcolor=blue,
    urlcolor=blue,
}

\begin{document}
\title{
Electronic Origin of Non-Zone-Center Phonon Condensations: \\ Octahedral Rotations as A Case Study
}
\author{Suguru Yoshida}
\email[e-mail:]{suguru.yoshida0224@gmail.com}
\affiliation{Department of Applied Chemistry, Kyushu University, Motooka, Fukuoka 819-0395, Japan}
\author{Hirofumi Akamatsu}
\affiliation{Department of Applied Chemistry, Kyushu University, Motooka, Fukuoka 819-0395, Japan}
\author{Katsuro Hayashi}
\affiliation{Department of Applied Chemistry, Kyushu University, Motooka, Fukuoka 819-0395, Japan}

\begin{abstract}
   Unstable zone-boundary phonon modes drive atomic displacements linked to a rich array of properties. 
   Yet, the electronic origin of the instability remains to be clearly explained.
   Here, we propose that bonding interaction between Bloch states belonging to different wavevectors leads to such instability
   via the pseudo- or second-order Jahn--Teller effect.
   Our first-principles calculations and representation theory-based analyses show that rotations of anion coordinated octahedra,
   an archetypal example of zone-boundary phonon condensations, are induced by this bonding mechanism.
   The proposed mechanism is universal to any non-zone-center phonon condensations and could offer a general approach to understand the origin of structural phase transitions in crystals.
\end{abstract}
\maketitle

Collective changes in atom positions often have drastic influences on the physical properties of solids.
Examples include polar atomic displacements emerging from the condensation of a transverse optic phonon at Brillouin-zone center (wavevector $\bm{q} = 0$)~\cite{n1}
through which the crystal acquires ferroelectric nature.
Displacements with finite wavevectors can also lead to functionalities inaccessible by zone-center phonon condensations.
Particularly, rotations of rigid polyhedral units, which arise from the condensation of phonon modes at Brillouin-zone boundary,
have recently attracted attention due to their strong coupling to magnetic~\cite{kanamori1959JoPaCoS,goodenough1955PR,goto2004PRL}, electric~\cite{fennie2005PRB,akamatsu2014PRL,lu2020MA},
thermal~\cite{evans1996CM,senn2015PRL,bird2020PRB}, conducting~\cite{ferretti1965JoPaCoS,lu2015SR}, and emitting~\cite{hanzawa2019JACS} properties
as well as their utility for realizing cross-coupled multiferroics~\cite{gopalan2011NM,benedek2011PRL,pitcher2015S}.

Key electronic features that favor the zero-$\bm{q$} displacements have been unveiled for the past few decades~\cite{bersuker1966PL,cohen1992N,walsh2011CSR,benedek2013JPCC}, facilitating the design of new polar materials.
In contrast, up to now, available information on the mechanism driving nonzero-$\bm{q}$ displacements is almost limited to the classical information~\cite{n2}.
For example, octahedral rotations (ORs) in perovskite-like compounds have long been attributed to ionic size mismatch, i.e., a coordination preference of $A$-site cations~\cite{woodward1997ACSB}.
Even though this explanation is useful for foreseeing whether a given material will exhibit ORs or not, it lacks the ability to predict the rotational pattern,
which is actually sensitive to electron correlations as reported in Ref.~\cite{hampel}.
A deep insight would be highly desirable for harnessing nonzero-$\bm{q}$ displacements and 
paving a promising route to manipulate the materials functionality.

This manuscript proposes a quantum-mechanical and group-theoretical framework for explaining the driving force for nonzero-$\bm{q}$ displacements in terms of electronic band structures.  
We employ an approach adopting second-order Jahn--Teller (SOJT) effect~\cite{n3},
which has been successfully applied with point-group analyses to describe the electronic origin of zero-$\bm{q}$ displacements~\cite{bersuker1966PL} and molecular deformations~\cite{polinger1989,bersuker2013CR}.
Although previous studies have discussed mechanisms behind some zone-boundary distortion from the perspective of electron-lattice interactions~\cite{bersuker1973F,garcia-fernandez2010JPCL,mizoguchi2004JACS},
there are, to our knowledge, no reports offering a general approach applicable to any arbitrary displacement in a material.
Here, we utilize space-group representation theory to treat translational symmetry breaking by nonzero-$\bm{q}$ displacements 
and consequently demonstrate that the SOJT-based approach can go beyond zero-$\bm{q}$ displacements while preserving its general applicability. 
To give a practical example, we apply our method to \textit{A}-site-empty perovskites, i.e., ReO$_3$-type \textit{BX}$_3$ compounds
and reveal why the octahedra rotate even in the absence of anion-attracting \textit{A}-site cations.
Our findings highlight that the primary driving force for ORs is of electronic rather than geometric origin.

We start by reviewing perturbative treatments of the SOJT effect~\cite{rabe2007,rondinelli2009PRB}, which enables deriving selection rules that underlie our discussion.
Using perturbation theory, one can expand the total energy ($E$) of a system, whose Hamiltonian is $\mathcal{H}$, in terms of normal coordinate ($Q$)
about the equilibrium high-symmetry phase:
\begin{eqnarray}
  E \!\! &=& \!\! E_0 + \bra{0}\mathcal{H}^{(1)}\ket{0}Q \nonumber \\
        &+& \!\! \frac{1}{2} \!\! \left [  \bra{0}\mathcal{H}^{(2)}\ket{0} - 2 \sum_{n} \frac{|\bra{0}\mathcal{H}^{(1)}\ket{n}|^2}{E_n - E_0}\right ] \!\! Q^2 + \cdots,    
\label{eq:E-Q}
\end{eqnarray}
with
\begin{equation}
   \mathcal{H}^{(1)} = \left .\frac{\partial \mathcal{H}}{\partial Q}\right |_{Q=0}, \text{ and } 
   \mathcal{H}^{(2)} = \left .\frac{\partial^2 \mathcal{H}}{\partial Q^2}\right |_{Q=0}.
\end{equation}
$E_0$ and $E_n$ refer to the energy of the ground state $\ket{0}$ and excited state $\ket{n}$, respectively,
both of which are eigenstates of the Hamiltonian for the high-symmetry phase with the space group $\mathcal{G}$.
Let $\ket{0}$ and $\ket{n}$ transform as irreducible representations (irreps) $\Phi_0$ and $\Phi_n$ of $\mathcal{G}$, respectively. 
Note that $\mathcal{H}^{(1)}$ transforms as the same irrep as $Q$ and the corresponding phonon mode~\cite{landau1981}; it will be denoted by $\Phi_{\mathrm{P}}$.
Of the quadratic terms in $Q$, the first one is always positive favoring the high-symmetry structure ($Q = 0$).
On the other hand, the second term $- 2 \sum_{n} \frac{|\bra{0}\mathcal{H}^{(1)}\ket{n}|^2}{E_n - E_0}$ is negative unless the matrix element $\bra{0}\mathcal{H}^{(1)}\ket{n}$ is forced to vanish by symmetry.
Following two conditions should be fulfilled for the magnitude of the second term to be larger than that of the first one so that the system undergoes the energy-lowering structural distortion.
First, the direct product $\Phi_0 \otimes \Phi_\mathrm{P} \otimes \Phi_n$ should contain the totally symmetric representation of $\mathcal{G}$
or equivalently $\Phi_0 \otimes \Phi_n$ should contain $\Phi_\mathrm{P}$
to attain a nonzero value for $\bra{0}\mathcal{H}^{(1)}\ket{n}$~\cite{dresselhaus2008}, representing the mixing of two electronic states in response to displacement perturbation.
Second, the energy gap $E_n - E_0$ in the denominator should be small.
Therefore, a distortion occurs if the corresponding phonon mode is symmetry-allowed to invoke mixing of the ground and low-lying excited states.

\begin{figure}
   \includegraphics[width=8.6cm]{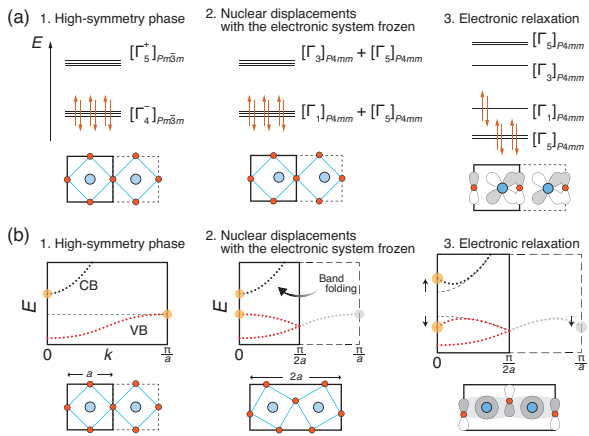}
   \caption{
   Step-by-step illustration for the SOJT mechanism driving (a) zone-center ($\bm{q}=0$) and (b) zone-boundary ($\bm{q} = \frac{\pi}{a}$) phonon condensation.
   The solid lines indicate the unit cells.
   }
\label{fig:illustration}
\end{figure}

The mixing of electronic states is referred to as covalent bond formations in the low-symmetry phase~\cite{bersuker2013CR}. 
Let us consider a polar displacement ($\bm{q} = 0$) in BaTiO$_3$ [Fig.~\ref{fig:illustration}(a)].
In the high-symmetry $Pm\bar{3}m$ structure, the valence band (VB, dominated by O 2\textit{p}) and conduction band (CB, dominated by Ti 3\textit{d}-$t_{\mathrm{2g}}$) have different symmetries 
belonging to distinct irreps, $\Gamma_{4}^-$ and $\Gamma_{5}^+$, respectively,
whereby the overlap of the two wavefunctions is forbidden~(step-1).
Once the Ti nuclei shift in a polar fashion ($\Gamma_4^-$) from their equilibrium positions, the crystal symmetry lowers from $Pm\bar{3}m$ to $P4mm$,
and concomitantly the degeneracies of the two electronic states are lifted~(step-2). 
How they split is defined by compatibility relations~\cite{dresselhaus2008}.
Two-thirds of O 2\textit{p} and Ti 3\textit{d}-$t_{\mathrm{2g}}$ states now transform as the same irrep $\left[ \Gamma_5 \right]_{P4mm}$~\cite{n4}
so that bonding and anti-bonding states appear through electronic relaxation, i.e., orbital mixing~(step-3).
Given the appropriate electron count, the bonding and non-bonding states accommodate the electrons to stabilize the low-symmetry configuration.
This is a chemistry explanation for the SOJT effect.

The perturbative and symmetry arguments have no restriction on the $\bm{q}$ value, implying that this framework should provide insights into the origin of nonzero-$\bm{q}$ displacements as well.
However, symmetry forbids any $\Gamma$--$\Gamma$ bond formation---like that shown in Fig.~\ref{fig:illustration}(a)---stemming from displacements with finite $\bm{q}$. 
Although this fact seems to rule out the possibility that the bonding mechanism is at play, nonzero-$\bm{q}$ displacements can give rise to, as we demonstrate later, mixing between two states with different $k$-vectors from each other
and thus arise through the SOJT mechanism.

Here we utilize the band structure to treat all Bloch states in a crystal, not just those at the $\Gamma$ point included in the energy diagram.
Figure~\ref{fig:illustration}(b) illustrares how a displacement ($\bm{q} = \frac{\pi}{a}$) results in the mixing of a VB state at $\bm{k} = \frac{\pi}{a}$ with a CB state at $\bm{k} = 0$,
where $a$ is the lattice constant of the high-symmetry phase.
The VB state cannot interact with the CB state in the high-symmetry configuration because of the discrepancy in $k$-vectors~(step-1).
Once the crystal experiences a distorting perturbation with the periodicity of $2a$, 
the unit cell doubles while folding the electronic bands into the halved first Brillouin zone~(step-2).
This band folding places the VB and CB states at the identical $k$-point, accepting the bond formation required to stabilize the distorted configuration. 
Note that the $k$-matching between two states under perturbation is merely a necessary condition for the states to mix;
namely, we must further examine whether the direct product $\Phi_0 \otimes \Phi_\mathrm{P} \otimes \Phi_n$ of the space-group irreps comprises the totally symmetric representation or not
in order to make sure that the distorting perturbation is symmetry-allowed to induce bond formation~(step-3).

By choosing ORs in ReO$_3$-type \textit{BX}$_3$ compounds as a case study, we illustrate how the SOJT-based approach integrated with space-group representation theory explains the emergence of nonzero-$\bm{q}$ displacements.
It is widely believed that, in perovskites, octahedra rotate to optimize the coordination environment for \textit{A}-site cations otherwise underbonded.
However, ReO$_3$-type compounds generally exhibit ORs rather than remain the aristotype $Pm\bar{3}m$ structure despite no \textit{A}-site cations; 
a majority of fluoride (pnictide) members crystallize in $R\bar{3}c$ ($Im\bar{3}$) structures involving out-of-phase $a^{-}a^{-}a^{-}$-type (in-phase $a^{+}a^{+}a^{+}$-type) ORs~\cite{glazer1972ACSB}. 
Hereafter, we also address this question as to what drives ORs.
First-principles calculations were performed for five ReO$_3$-type compounds---GaF$_3$, RhF$_3$, AlH$_3$, ReO$_3$, and RhP$_3$---and perovskite BaTiO$_3$ using the projector augmented-wave method~\cite{blochl1994PRB,kresse1999PRB}
and the HSE06 hybrid functional~\cite{heyd2003JCP,heyd2006JCP,krukau2006JCP} as implemented in \textsc{vasp} code~\cite{kresse1993PRBa,kresse1993PRB,kresse1996PRB,kresse1996CMS}.
Details are given in Supplemental Material~\cite{supp}.

\begin{figure}
   \includegraphics[width=8.6cm]{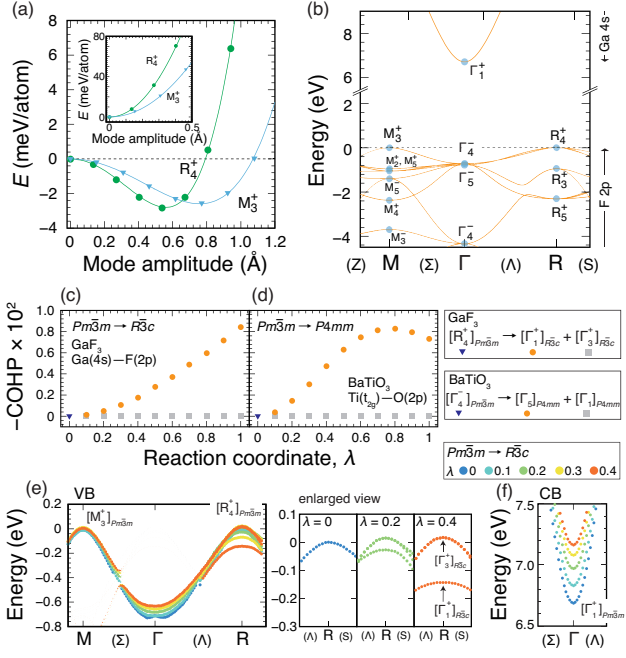}
   \caption{
   (a) Total energy and Madelung energy (shown in the inset) of GaF$_3$ as a function of the amplitude of out-of-phase ($\mathrm{R}_4^+$) and in-phase ($\mathrm{M}_3^+$) OR modes.
   (b) Electronic band structure for GaF$_3$ adopting $Pm\bar{3}m$ symmetry with the irrep labels identified by \textsc{irvsp} code~\cite{gao}.
   The band-resolved $-$COHPs for the (c) $\mathrm{R}_4^+$ VB states of GaF$_3$ and (d) $\Gamma_4^-$ VB states of BaTiO$_3$ as a function of reaction coordinate $\lambda$.
   The legend on the right explains how the VB states split as $\lambda$ rises from zero. 
   Band structures of GaF$_3$ in the energy region near the (e) VB maximum and (f) CB minimum as a function of $\lambda$,
   where all the bands including those of $R\bar{3}c$ structures ($\lambda \neq 0$) are drawn along the high-symmetry path of the $Pm\bar{3}m$ structure ($\lambda = 0$)
   via band-unfolding process with \textsc{PyProcar} code~\cite{herath2020CPC}.
   }
\label{fig:GaF3}
\end{figure}

We first focus on GaF$_3$, whose $R\bar{3}c$ structure has been experimentally identified~\cite{roosand2001ZFK-NCS}.
Figure~\ref{fig:GaF3}(a) shows that substantial energy gains relative to the $Pm\bar{3}m$ configuration are observed for $a^{-}a^{-}a^{-}$-type and $a^{+}a^{+}a^{+}$-type OR modes 
transforming as the irrep $\mathrm{R}_4^+$ and $\mathrm{M}_3^+$, respectively.
It means that the ORs are energetically favorable at the level of density functional theory.
On the other hand, ORs are unfavorable at the level of an electrostatic model, as evident from the inset of Fig.~\ref{fig:GaF3}(a).
This qualitative disagreement corroborates that non-classical behavior of electrons excluded in Madelung energy calculations is vital for understanding stabilization mechanisms behind ORs.
In the following, we consider how the $R\bar{3}c$ phase appears through the SOJT mechanism, i.e., the case of $\Phi_{\mathrm{P}} = \mathrm{R}_4^+$. 
Discussions on the in-phase $Im\bar{3}$ phase and its difference from the $R\bar{3}c$ phase in terms of orbital interactions are given in Ref.~\cite{supp}.

The calclated electronic band structure [Fig.~\ref{fig:GaF3}(b)] shows that the VB maximum (CB minimum) of GaF$_3$ with the $Pm\bar{3}m$ structure is at R ($\Gamma$) point
and transforms like the irrep $\mathrm{R}_4^+$ ($\Gamma_1^+$) of $Pm\bar{3}m$.
Using the irreps of the VB maximum, CB minimum, and considered distortion mode, we calculate the direct product $\Phi_0 \otimes \Phi_\mathrm{P} \otimes \Phi_n$
with \textsc{dirpro} tool~\cite{aroyo2006ACA}:
\begin{equation}
   \mathrm{R}_4^+ \otimes \mathrm{R}_4^+ \otimes \Gamma_1^+ = \Gamma_1^+ + \Gamma_3^+ + \Gamma_4^+ + \Gamma_5^+.
\label{eq:dirpro}
\end{equation}
The result contains $\Gamma_1^+$, i.e., the totally symmetric representation of $Pm\bar{3}m$ so that the selection rule is fulfilled,
allowing the $\mathrm{R}_4^+$ distortion to give rise to an $\mathrm{R}_4^+$--$\Gamma_1^+$ interaction.
In other words, the $a^-a^-a^-$-type ORs are likely to be attributable to the SOJT mechanism.
Compatibility relations obtained with \textsc{correl} tool~\cite{aroyo2006ACA} reveal that the VB and CB extrema of the $Pm\bar{3}m$ phase respectively split as follows:
\begin{equation}
   \left[ \mathrm{R}_4^+ \right]_{Pm\bar{3}m} \rightarrow \left[\Gamma_1^+ \right]_{R\bar{3}c} + \left[\Gamma_3^+ \right]_{R\bar{3}c}, \\ 
\label{eq:comp_R}
\end{equation}
and
\begin{equation}
\label{eq:comp_GM}
   \left[ \Gamma_1^+ \right]_{Pm\bar{3}m} \rightarrow \left[ \Gamma_1^+ \right]_{R\bar{3}c}.
\end{equation}
One can expect that the new bonding and anti-bonding states belong to the irrep $\left[\Gamma_1^+\right]_{R\bar{3}c}$ and
that the $\left[\Gamma_3^+\right]_{R\bar{3}c}$ state remains non-bonding due to the absence of CB states with the same symmetry.
Although some $\mathrm{X}$--$\mathrm{M}$ interactions couple to the $\mathrm{R}_4^+$ distortion~\cite{supp},
we ignore them here as their large energy gaps.

We next assess the dependence of band-resolved projected crystal orbital Hamiltonian population (COHP)~\cite{dronskowski1993JPC,deringer2011JPCA,maintz2013JCC,sun2019JCC}
between F 2\textit{p} and Ga 4\textit{s} states---composing the $\mathrm{R}_4^+$ and $\Gamma_1^+$ states, respectively~\cite{supp}---on the
rotation magnitude.
The rotation distortion is parametrized by reaction coordinate $\lambda$ varying from 0 (fully relaxed high-symmetry structure) to 1 (fully relaxed low-symmetry structure).
We use negative-signed COHPs ($-$COHPs) whose positive (negative) values represent bonding (anti-bonding) interactions.
Figure~\ref{fig:GaF3}(c) illustrates that the $-$COHPs for the $\mathrm{R}_4^+$ VB states change significantly with $\lambda$.
While there is no bonding interaction for the VB states in the $Pm\bar{3}m$ configuration ($\lambda = 0$),
these states split at finite $\lambda$, and the $-$COHP for the $\left[\Gamma_1^+ \right]_{R\bar{3}c}$ state increases on approaching $\lambda = 1$.
This behavior proves that the rotation magnitude strongly correlates with the degree of $\mathrm{R}_4^+$--$\Gamma_1^+$ bonding interaction.
Note that the $-$COHP for the $\left[\Gamma_3^+ \right]_{R\bar{3}c}$ states remains close to zero, supporting the non-bonding nature expected from the symmetry arguments. 
By checking the decrease in the $-$COHP with increasing $\lambda$,
we also validate the anti-bonding character of $\Gamma_1^+$ CB state with which the considered VB states mix~\cite{supp}. 
Figure~\ref{fig:GaF3}(d) plots the $-$COHPs between Ti $3d$-$t_{\mathrm{2g}}$ and O 2\textit{p} states for the $\Gamma_4^-$ VB states of BaTiO$_3$ as a function of $\lambda$,
where the low-symmetry structure corresponds to $P4mm$ one.
Comparing Figs.~\ref{fig:GaF3}(c) and \ref{fig:GaF3}(d) reveals that the evolutions of bonding interactions in GaF$_3$ are very similar to those in BaTiO$_3$,
implying that Ga(4\textit{s})--F(2\textit{p}) bonding in GaF$_3$ drives the ORs in the same way that Ti(3\textit{d})--O(2\textit{p}) bonding drives the polar displacements in BaTiO$_3$.

The covalent bonds in BaTiO$_3$ cause a shift down (up) in energy of its occupied bonding (unoccupied anti-bonding) state to produce a net energy gain to the polar phase~\cite{filippetti2002PRB}.
Here, we demonstrate that by calculating band dispersions with varying $\lambda$, a similar stabilization arises from $\mathrm{R}_4^+$--$\Gamma_1^+$ bonding interactions accompanied by the $a^-a^-a^-$-type ORs.
In the $R\bar{3}c$ structures where the $\mathrm{R}_4^+$ VB states split [Eq.~(\ref{eq:comp_R})], 
the increase in $\lambda$ (and therefore in the OR magnitude) lowers the energy of the bonding $\left[\Gamma_1^+ \right]_{R\bar{3}c}$ state while
keeping that of the non-bonding $\left[\Gamma_3^+ \right]_{R\bar{3}c}$ state almost unchanged [Fig.~\ref{fig:GaF3}(e)].
Also, destabilization of the CB $\Gamma_1^+$ state in response to the ORs is manifest in Fig.~\ref{fig:GaF3}(f).
Thus, we find that the ORs---unfavorable in terms of Madelung energy---become energetically favorable due to the bonding interaction that stabilizes the electronic system.
There is no conceptual difference from the polar shifts in BaTiO$_3$, except that for the case of ORs the interacting Bloch states locate at distinct $k$-points in the high-symmetry configuration.
Note in Fig.~\ref{fig:GaF3}(e) that the energy of the $\mathrm{M}_3^+$ VB states is quite insensitive to the $a^-a^-a^-$-type ORs,
as expected from symmetry considerations~\cite{supp}.
Instead, the increase in the $a^+a^+a^+$-type rotation in magnitude lowers the energy of the $\mathrm{M}_3^+$ state but does not influence the $\mathrm{R}_4^+$ state~\cite{supp}.

\begin{figure}
   \includegraphics[width=8.6cm]{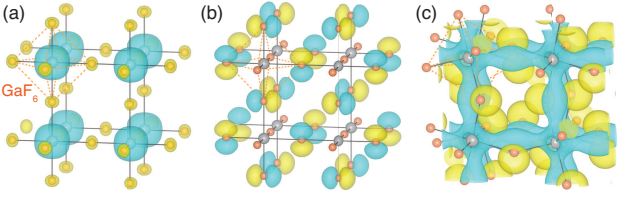}
   \caption{
   Real part of the wavefunctions for (a) $\left[\Gamma_1^+ \right]_{Pm\bar{3}m}$ CB and (b) $\left[\mathrm{R}_4^+ \right]_{Pm\bar{3}m}$ VB states of the fully relaxed $Pm\bar{3}m$ structure,
   and for (c) $\left[\Gamma_1^+ \right]_{R\bar{3}c}$ VB state of the fully relaxed $R\bar{3}c$ structure. 
   They are extracted using \textsc{vaspkit}~\cite{wang} and visualized by \textsc{vesta}~\cite{momma2011JAC}.
   Yellow and blue isosurfaces denote positive and negative lobes, respectively.
   }
\label{fig:wf}
\end{figure}

\begin{figure}
   \includegraphics[width=8.6cm]{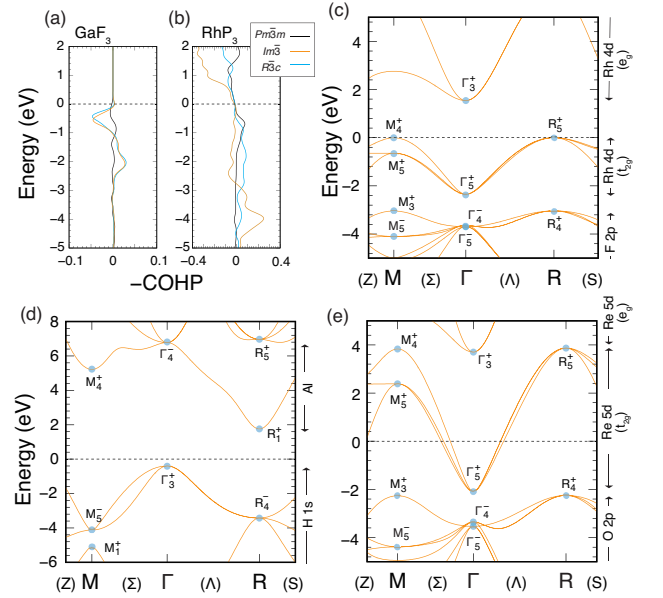}
   \caption{
   Averaged $-$COHPs of anion--anion bondings~\cite{supp} in the $Pm\bar{3}m$, $R\bar{3}c$, and $Im\bar{3}$ structures of (a) GaF$_3$ and (b) RhP$_3$.
   Electronic band structures for (c) RhF$_3$, (d) AlH$_3$, and (e) ReO$_3$ with $Pm\bar{3}m$ structures along with the irrep labels.
   }
\label{fig:band}
\end{figure}

Next, we provide a real-space picture of the bond formation in GaF$_3$.
The $Pm\bar{3}m$ configuration having linear Ga--F--Ga chains results in an equal amount of constructive and destructive
overlap between the $\Gamma_1^+$ CB and $\mathrm{R}_4^+$ VB states; the two are orthogonal [Figs.~\ref{fig:wf}(a) and \ref{fig:wf}(b)].
When the ORs occur so as to bend Ga--F--Ga angle, however, the two states are no longer orthogonal and can mix to form a low-energy bonding state.
Indeed, the $\left[\Gamma_1^+ \right]_{R\bar{3}c}$ VB state now exhibits a substantial wavefunction's magnitude in an area
between the Ga and F sites~[Fig.~\ref{fig:wf}(c)].
This \textit{sp}$\pi$ bonding is reminiscent to the \textit{dp}$\pi$ and \textit{dp}$\sigma$ bondings of BaTiO$_3$~\cite{cohen1992N,hickox-young2020PRB}.
One might expect anion--anion bonds to stabilize the tilted structure because such a stabilization mechanism is well established in skutterudites like RhP$_3$~\cite{lefebvre-devos2001PRB,luo2015NC,hanus2017CM},
whereas both the F--F bonding and anti-bonding states appear below the Fermi energy and offer no net stabilizing effect~[Fig.~\ref{fig:band}(a)].
This is in striking contrast to P--P bonds~[Fig.~\ref{fig:band}(b)], i.e., P$_4$ polyanionic rings in RhP$_3$ 
due to which the $Im\bar{3}$ structure is substantially lower in energy relative to $R\bar{3}c$ and $Pm\bar{3}m$ phases~\cite{supp}.

Both RhF$_3$ and AlH$_3$ crystallize in $R\bar{3}c$ structures~\cite{grosse1987ZFAAC,turley1969IC}, where band dispersions near the Fermi levels (and hence the symmetry of wavefunctions) are considerably different from those of GaF$_3$.
The CB of RhF$_3$ mainly consists of 4\textit{d} rather than 4\textit{s} states [Figs.~\ref{fig:band}(c)],
and the VB of AlH$_3$ is dominated by H 1\textit{s} states instead of 2\textit{p} states of O or F [Fig.~\ref{fig:band}(d)].
Despite such differences, our direct product calculations prove that the $\mathrm{R}_4^+$ distortion can be stabilized in RhF$_3$ and AlH$_3$ by  
$\mathrm{R}_5^+$--$\Gamma_3^+$ and $\mathrm{R}_4^-$--$\Gamma_4^-$ bonding, respectively~\cite{supp}.
Also in ReO$_3$, the pair of $\mathrm{R}_5^+$ and $\Gamma_3^+$ states is symmetry-allowed to interact under the distortion and results in bonding and anti-bonding orbitals.
However, both of them are unoccupied, producing no net energy gain [Fig.~\ref{fig:band}(e)].
ReO$_3$ is therefore predicted to retain the aristotype $Pm\bar{3}m$ structure, consistent with experimental reports~\cite{meisel1932ZFAAC, bozin2012PRB}.

Generally, more than one pair of Bloch states around the Fermi level interact under a given distortion.
For example in ReO$_3$, $\mathrm{R}_4^+$ displacive perturbation permits $\mathrm{R}_4^+$--$\Gamma_3^+$ and $\mathrm{R}_4^+$--$\Gamma_5^+$ interactions as well~\cite{supp}.
Although a stabilizing effect is expected from the former interaction, this would be counteracted by an energy penalty due to the populated anti-bonding state arising from the latter. 
We believe that calculating $\bra{0}\mathcal{H}^{(2)}\ket{0}$ and $\bra{0}\mathcal{H}^{(1)}\ket{n}$'s separately enables quantitative discussions on such competing effects
through decoupling the contributions that repulsions and hybridizations play in determining the sign of the quadratic coefficient of Eq.~(\ref{eq:E-Q});
it may need further methodological developments.
Such calculations may also allow incorporating the effect of hybridization into the tolerance factor approach~\cite{goldschmidt1926N}, leading to a new descriptor for the structural instability.

To summarize, we have proposed that the SOJT effect, when combined with the band folding scenario, can be utilized to uncover the driving mechanism of nonzero-$\bm{q}$ displacements.
Based on this idea, we have demonstrated that energy-lowering \textit{B}--\textit{X} bondings trigger the ORs even with the empty \textit{A}-site cavities. 
The \textit{A}-site cations would play a secondary role in perovskites.
While only zone-boundary distortions are discussed here, the same framework can obviously apply to any distortions including incommensurate modulations.
We hope this study leads to a unified description of a variety of structural distortions in solids that will be exploited for rational property design.

\begin{acknowledgments}
This work was supported by Research Fellowships of Japan Society of the Promotion of Science (JSPS) for Young Scientists, 
the Grant-in-Aid for JSPS Research Fellow (Grant Number JP20J01149),
and JSPS KAKENHI (Grant Numbers JP18H01892 and JP16H06440). 
The computation was carried out using the computer resource offered under the category of General Projects
by Research Institute for Information Technology, Kyushu University.
\end{acknowledgments}


%

\begin{thebibliography}{75}
\bibitem [{n1()}]{n1}%
  \BibitemOpen
  \href@noop {} {}\bibinfo {note} {We use the symbol $\bm{q}$ and $\bm{k}$ to
  denote wavevectors of phonons (and distortions) and electronic states,
  respectively.}\BibitemShut {Stop}%
\bibitem [{n2()}]{n2}%
  \BibitemOpen
  \href@noop {} {}\bibinfo {note} {One of the exceptions is Peierls transition,
  where a one-dimensional crystal with Fermi wavevector $k_{\mathrm{F}}$
  undergoes a distortion with $q = 2k_{\mathrm{F}}$.}\BibitemShut {Stop}%
\bibitem [{n3()}]{n3}%
  \BibitemOpen
  \href@noop {} {}\bibinfo {note} {We follow the terminology in the community
  of ferroics and use the term ``second-order'' rather than
  ``pseudo''}\BibitemShut {NoStop}%
\bibitem [{n4()}]{n4}%
  \BibitemOpen
  \href@noop {} {}\bibinfo {note} {To prevent ambiguity, we sometimes use the
  subscript symbol denoting the space group with respect to which the irrep is
  defined.}\BibitemShut {Stop}%
\end{thebibliography}
\end{document}